# Proof of the quantum null energy condition for free fermionic field theories

Taha A. Malik[*] and Rafael Lopez-Mobilia[†]

*Department of Physics and Astronomy, The University of Texas at San Antonio, San Antonio, Texas 78249, USA*

The quantum null energy condition (QNEC) is a quantum generalization of the null energy condition, which gives a lower bound on the null energy in terms of the second derivative of the von Neumann entropy or entanglement entropy of some region with respect to a null direction. The QNEC states that $\langle T_{kk} \rangle_p \geq \lim_{A \to 0}(\frac{\hbar}{2\pi A} S''_{\text{out}})$, where $S_{\text{out}}$ is the entanglement entropy restricted to one side of a codimension-2 surface $\Sigma$, which is deformed in the null direction about a neighborhood of point $p$ with area $A$. A proof of QNEC has been given before, which applies to free and super-renormalizable bosonic field theories, and to any points that lie on a stationary null surface. Using similar assumptions and methods, we prove the QNEC for fermionic field theories.

## I. INTRODUCTION

In general relativity, energy conditions are restrictions imposed on the energy-momentum tensor of matter and (nongravitational) fields to prevent unphysical solutions of Einstein's field equations. There are several energy conditions, of which the null energy condition (NEC) is one of the weakest. It states that

$$T_{kk} = T_{ab} k^a k^b \geq 0, \quad (1)$$

where $T_{ab}$ is the energy-momentum tensor and $k_a$ is an arbitrary null vector. In spite of being weaker than other energy conditions, the NEC is sufficient to prove many important results, such as the Penrose singularity theorem [1], the second law of black hole thermodynamics [2], and other area laws [3] etc.

It is well known, on the other hand, that the NEC and all other energy conditions are violated by quantum field theories, even free ones. So finding generalizations of these conditions when quantum fields are included is an important issue which could help generalize other results in general relativity that depend on classical energy conditions, provide insights into the nature of quantum gravity, and further highlight the connection between energy and information. For example, the vacuum modular Hamiltonian of a Rindler wedge is given by the boost generator for any quantum field theory [4]. The QNEC has been used to investigate modular Hamiltonian for more general half spaces [5].

The QNEC was conjectured by considering the covariant entropy bound [6–8], an entropy bound which attempts to reformulate and generalize the Bekenstein entropy bound [9,10] in a covariant way, and the related quantum focusing conjecture [11], a generalization of the focusing theorem for null geodesics. The QNEC states that

$$\langle T_{kk} \rangle_p \geq \lim_{A \to 0} \left[ \frac{\hbar}{2\pi A} S''_{\text{out}}(\Sigma) \right], \quad (2)$$

where $S_{\text{out}}$ is the von Neumann entropy or entanglement entropy restricted to one side of a codimension-2 surface $\Sigma$ with the derivatives taken with respect to deformations in the null direction about a neighborhood of point $p$ with infinitesimal transverse area $A$. More details are provided in Sec. II.

The QNEC has been proved for holographic theories and CFTs with a twist gap [12–15]. In this paper, we provide a new proof of the QNEC for free and super-renormalizable fermionic field theories using a similar method for the proof with bosonic field theories [16].

### A. Overview

The QNEC has been proved for free and super-renormalizable bosonic field theories at any point that lies on a stationary null surface $N$, with $\Sigma$ being an arbitrary cut and deformed along $N$ [16] (see Fig. 1). The restriction to such points was necessary as the proof relied on quantization on a null surface (null quantization), which can only be applied to stationary null surfaces. Such null surfaces include the Rindler horizon in Minkowski space and the horizon of eternal Schwarzschild and Kerr black holes.

[*]taha.malik11@alumni.imperial.ac.uk, taha.malik@utsa.edu
[†]rafael.lopezmobilia@utsa.edu

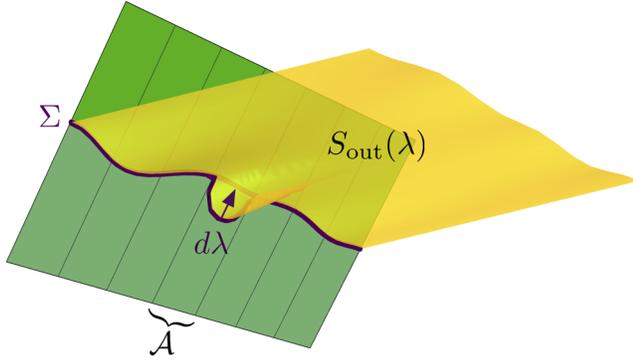

FIG. 1. The codimension-2 surface Σ splits some Cauchy surface into two disjoint regions. One side (yellow surface) is unitarily equivalent to the stationary null surface $N$ (green surface) to the future of Σ with part of the future null infinity. Hence, $S_{\text{out}}$ on both these surfaces are equal. Deformations of Σ are equivalent to deformations on the distinguished pencil. [Reprinted figure with permission from R. Bousso, Z. Fisher, J. Koeller, S. Leichenauer, and A. C. Wall, Phys. Rev. D **93**, 024017 (2016). Copyright 2016 by the American Physical Society.]

Null quantization allows the proof to reduce to just working with decoupled free left-moving chiral fermionic $1+1$ CFTs and an auxiliary system whose exact details are not needed. The state of the system was constructed via a path integral and various traces evaluated on replicated manifolds to calculate the Renyi entropies. The replica trick was used to calculate the von Neumann entropy via analytically continuing the Renyi entropies, which allowed the explicit evaluation of $\frac{\hbar}{2\pi A} S''_{\text{out}} - \langle T_{kk} \rangle$ to zeroth order in $A$. To prove the QNEC, it is sufficient to prove that the expression above is less than or equal to zero for small $A$. This will also be our strategy for proving the QNEC for free and super-renormalizable fermionic field theories.

## II. THE QUANTUM NULL ENERGY CONDITION

If we choose a point $p$ in a globally hyperbolic spacetime, a null vector $k^a$, and any smooth codimension-2 surface $\Sigma$[1] that

(i) Σ partitions a Cauchy surface into two,
(ii) Its boundary $\partial\Sigma$ contains the point $p$, and
(iii) Σ is normal to $k^a$ with vanishing expansion at $p$,

then we can consider deforming the surface Σ about a neighborhood of $p$, with a small area $A$ along the null geodesics generated by $k^a$. By choosing an affine parameter $\lambda$ for the null generators, we can label the deformed surfaces as $\Sigma(\lambda)$ with $\Sigma(0) = \Sigma$. Then the QNEC states that

$$\langle T_{kk}\rangle_p \geq \lim_{A\to 0}\left\{\frac{\hbar}{2\pi A} S''_{\text{out}}[\Sigma(\lambda)]|_{\lambda=0}\right\}, \quad (3)$$

where $S_{\text{out}}$ is the entanglement entropy on one side of $\Sigma(\lambda)$ and the derivative is taken with respect to $\lambda$. The choice of side can be done arbitrarily.

---
[1]There are an infinite number of choices for Σ.

To be more explicit, we can set up a transverse coordinate system ($y$ coordinate) for $\partial\Sigma$ in a neighborhood of $p$ and define null vectors $k^a(y)$ normal to $\partial\Sigma$ with $k^a(p) = k^a$. The vanishing expansion condition is with respect to this family of null vectors.

With our choice of affine parameter, we now have a coordinate system $(\lambda, y)$ with the location of $\partial\Sigma$ at $(0, y)$. Deformations of Σ now correspond to its boundary $\partial\Sigma$ shifting along the $\lambda$ coordinate about $p$. $\Sigma(\lambda)$ is located at $[V(y,\lambda), y]$ where $V(y,\lambda) = 0$ everywhere except in a neighborhood of $p$ with transverse area $A$ for which $V(y,\lambda) = \lambda$.

Our proof of the QNEC applies to free and super-renormalizable fermionic field theories for which $p$ lies on a stationary null surface (vanishing null expansion everywhere) with a normal vector $k^a$ and Σ a section of this null surface, as in the bosonic case [16]. Without a loss of generality, we choose $k^a$ to be future directed and choose the side of Σ towards which $k^a$ points. Initially, we work in dimension $D > 2$ to ensure we have a transverse direction to deform Σ on.

## III. NULL DISCRETIZATION AND QUANTIZATION

A stationary null surface $N$ can be obtained as a limit of Cauchy surfaces and by unitary invariance; $S_{\text{out}}$ on the Cauchy surfaces and the null surface $N$ with part of the future null infinity are equal. Hence, by using the null surface as an initial surface and quantizing on $N$ (null quantization), we can calculate $S_{\text{out}}$ by restricting the state to the future of Σ. Null quantization requires $N$ to be a stationary null surface [17,18].

We first discretize $N$ along the transverse direction ($y$ direction) into regions of small transverse area $A$ called pencils. We take this transverse area to be the same as the area in Eq. (3) (i.e., the size of the neighborhood of $p$ about which Σ is deformed) as we take the limit when $A \to 0$. In this way, deformations of Σ are equivalent to deformations along the distinguished pencil (the pencil which contains $p$).

With this setup, it has been shown [16,17] that restricted to $N$, the theory decomposes into a product of $1 + 1$ dimensional free left-moving chiral conformal field theory (CFT), with $\frac{K}{2}$ CFTs associated to each pencil of $N$, where $K$ is the number of components of the spinor field, and thus, the vacuum state factorizes with respect to this pencil decomposition of $N$. For small $A$, the state of the system can be written as [16]

$$\rho(\lambda) = \rho_{\text{pen}}^{(0)}(\lambda) \otimes \rho_{\text{aux}}^{(0)} + \sigma(\lambda), \qquad (4)$$

where $\rho_{\text{pen}}^{(0)}(\lambda)$ is the vacuum state on the distinguished pencil restricted to the future of $\Sigma(\lambda)$, $\rho_{\text{aux}}^{(0)}$ is some state on all the other pencils on $N$, and part of future null infinity (auxiliary system) and $\sigma(\lambda)$ is a small perturbation. For the proof of the QNEC, we only need to consider terms in $\sigma(\lambda)$ up to order $A^{\frac{1}{2}}$. Such terms are constructed by taking the partial trace of terms of the form $|0\rangle\langle 1|$ and $|1\rangle\langle 0|$ in the distinguished pencil Fock basis because they scale like $A^{\frac{1}{2}}$. In general, $|n\rangle\langle m|$ scales like $A^{\frac{n+m}{2}}$ [16]. This will be partially reviewed in Sec. IV. From now on, we refer to the distinguished pencil simply as the pencil and all other pencils together as the *auxiliary system*.

When "restricted" to the full pencil, the state is near the vacuum state $\rho_{\text{pen}}^{(0)}$ for small $A$. Up to the order $A^{\frac{1}{2}}$, the state on $N$ must be of the following form[2]:

$$\rho = |0\rangle\langle 0| \otimes \rho_{\text{aux}}^{(0)} + A^{\frac{1}{2}} \sum_{ij} (|0\rangle\langle \psi_{ij}^1| + |\psi_{ij}^2\rangle\langle 0|) \otimes |i\rangle\langle j|$$

$$+ A^{\frac{1}{2}} \sum_{ij} (|0\rangle\langle \psi_{ij}^2| + |\psi_{ij}^1\rangle\langle 0|) \otimes |j\rangle\langle i| + \ldots, \qquad (5)$$

where[3] $|i\rangle$ and $|j\rangle$ together form an orthonormal basis for the auxiliary system such that $|i\rangle\langle j|$ and $|j\rangle\langle i|$ together forms a Grassmann-odd basis of operators. For example, $|i\rangle$ could label states obtained by applying even creation operators to the vacuum state ($a_1^\dagger a_2^\dagger|0\rangle$) and $|j\rangle$ labels states obtained by applying odd creation operators to the vacuum state ($a_1^\dagger a_2^\dagger a_3^\dagger|0\rangle$). Without a loss of generality, we pick $|i\rangle$ and $|j\rangle$ that diagonalizes $\rho_{\text{aux}}^{(0)}$.

$|\psi_{ij}\rangle$ are single particles states in the CFT so that $|\psi_{ij}\rangle\langle 0|$ and $|0\rangle\langle\psi_{ij}|$ takes the schematic form $|0\rangle\langle 1|$ and $|1\rangle\langle 0|$ and represents the perturbation $\sigma$ of the state on $N$ at order $A^{\frac{1}{2}}$ (see Sec. III). Higher order terms in $A$ represented in "..." can be ignored.

The single particle states $|\psi_{ij}\rangle$ can be constructed applying the single-field operator on the vacuum state. In the Euclidean path integral picture, the most general single particle states can be constructed by insertions of the single-field in the Euclidean plane, as shown in Fig. 2. Since chiral fermionic fields only depend on the coordinate $z = x + t = x - i\tau$, translations along the Rindler horizon (null surface) are equivalent to translations along the spatial $x$ direction [16].

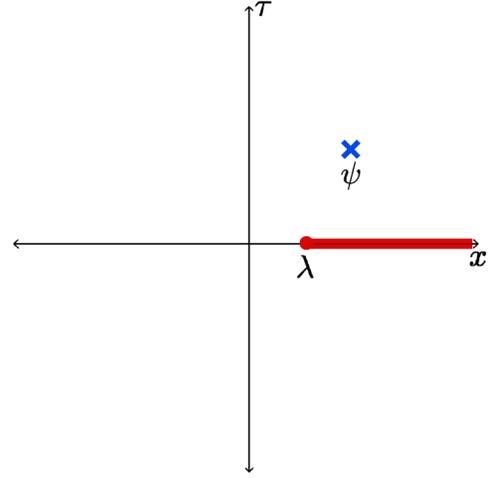

FIG. 2. The state of the CFT can be constructed in the path integral picture with insertions of $\psi$. The state on the full pencil is obtained by moving the branch point to $x = -\infty$. The state for some value $\lambda$ is obtained by moving the branch point to $x = \lambda$.

To trace out the degrees of freedom over $x < 0$ and obtain the state of the system when $\lambda = 0$, we insert a branch cut from the origin to $x = \infty$ on the Euclidean plane. The matrix elements of the state are represented via the path integral picture schematically[4] as [19]

$$\langle\psi|\rho|\psi'\rangle = M^{-1} \int_{\psi(x>0,0^-)=\psi'(x)}^{\psi(x>0,0^+)=-\psi(x)} [d\psi] O_\rho e^{-S_E}, \qquad (6)$$

with boundary conditions defined just above and below the branch cut such that $\hat{\psi}(x)|\psi\rangle = \psi(x)|\psi\rangle$. $M$ is a normalizing constant, and $O_\rho$ is some operator that defines $\rho$. In our case, $O_\rho$ will be of the form $I + A^{\frac{1}{2}}\tilde{O}_\rho$, with $\tilde{O}_\rho$ constructed from single field insertions.

To obtain the state of the system for any $\lambda$, we need to take the partial trace along $x < \lambda$ on the pencil. This can be done by moving the branch point from $x = 0$ to $x = \lambda$. Alternatively, we can translate the field insertions to the left by $\lambda$. From this point of view, the vacuum state,

$$\rho_{\text{pen}}^{(0)} = e^{-2\pi K_{\text{aux}}} \qquad (7)$$

---

[2] For simplicity, we have assumed that we have one chiral CFT on the pencil. To include more CFTs, one can add the corresponding terms to Eq. (5). However, since we only need to consider the second order term of the entropy (see Sec. IV), we can consider the extra terms separately and simply sum over them to obtain $S^{(2)}$.

[3] We have ensured that the state $\rho$ is a physical state in the sense that the state is Hermitian and invariant under a $2\pi$ rotation. For example, if we consider the vacuum state $|0\rangle$ and the single particle state $|1\rangle$ for some fermionic quantum system, then $\bar{\rho}_1 = |0\rangle\langle 0| + |1\rangle\langle 1|$ is a physical state but $\bar{\rho}_2 = |0\rangle\langle 1| + |0\rangle\langle 1|$ is an unphysical state since after a rotation by $2\pi$, $\bar{\rho}_2 \to -\bar{\rho}_2$. In other words, a physical state must be a Grassmann-even operator.

[4] For a fermionic system, the functional integral is in terms of Grassmann variables. Thus, when taking the trace in the functional integral, we have to sum over antidiagonal elements, and so the upper boundary condition has an extra minus sign. Additionally, one can not directly construct a path integral as in Eq. (6) for Majorana spinors without some modification. See the discussion in Sec. V for more details.

does not depend on $\lambda$ [16]. The modular Hamiltonian $K_{\text{pen}}$ is also the Rindler boost generator [20] (up to an additive constant).

Using this, we can evaluate the trace of the state [Eq. (5)] up to $x = \lambda$ as[5]

$$\rho(\lambda) = \rho_{\text{pen}}^{(0)} \otimes \rho_{\text{aux}}^{(0)} + \left[ A^{\frac{1}{2}} e^{-2\pi K_{\text{tot}}} \sum_{ij} \int dr d\theta f_{ij}(r,\theta) \psi(re^{i\theta} - \lambda) \otimes E_{ij}(\theta) \right] + \left[ A^{\frac{1}{2}} e^{-2\pi K_{\text{tot}}} \sum_{ij} \int dr d\theta f_{ji}(r,\theta) \psi(re^{i\theta} - \lambda) \otimes E_{ji}(\theta) \right], \quad (8)$$

where
(i) $K_{\text{tot}} = K_{\text{pen}} + K_{\text{pen}}$ with $K_{\text{aux}}$ defined so that $\rho_{\text{aux}}^{(0)} = e^{-2\pi K_{\text{aux}}}$,
(ii) $E_{ij}(\theta) := e^{\theta K_{\text{aux}}} |i\rangle\langle j| e^{-\theta K_{\text{aux}}} = e^{\theta(K_i - K_j)} |i\rangle\langle j|$, and similarly,
(iii) $E_{ji}(\theta) := e^{\theta(K_j - K_i)} |j\rangle\langle i|$.

Recall that we chose a basis for the auxiliary system such that $\rho_{\text{aux}}^{(0)}$ is diagonal. Hence, $K_{\text{aux}}$ is also diagonal with an eigenvalue $K_{\text{aux}}|i\rangle := K_i|i\rangle$. If we define $f_{ii}(r, \theta) = f_{jj}(r, \theta) = 0$ for all $i$ and $j$, then we can rewrite Eq. (8) as

$$\rho(\lambda) = \rho_{\text{pen}}^{(0)} \otimes \rho_{\text{aux}}^{(0)} + \left[ A^{\frac{1}{2}} e^{-2\pi K_{\text{tot}}} \right. \\ \left. \times \sum_{\mu,\nu} \int dr d\theta f_{\mu\nu}(r,\theta) \psi(re^{i\theta} - \lambda) \otimes E_{\mu\nu}(\theta) \right], \quad (9)$$

where $\mu, \nu$ are both summed over $i$ and $j$ indices. For definiteness, we define for all $\mu$ and $\nu$,

$$E_{\mu\nu}(\theta) := e^{\theta K_{\text{aux}}} |\mu\rangle\langle\nu| e^{-\theta K_{\text{aux}}} = e^{\theta(K_i - K_j)} |\mu\rangle\langle\nu|, \quad (10)$$

which agrees with the previous definition for matching $i$ and $j$.

To ensure that the state (9) is Hermitian, we must impose the following reality condition on $f_{\mu\nu}(r, \theta)$:

$$f_{\mu\nu}(r, \theta) = -i f^*_{\nu\mu}(r, 2\pi - \theta). \quad (11)$$

Other than this condition, $f_{\mu\nu}(r, \theta)$ can chosen arbitrarily (see the Appendix for more details).

---
[5]To see this, consider angular quantization with an origin at $(t, x) = (0, 0)$. Then, isolating the single field insertion, we have that $\langle\psi|\rho|\psi'\rangle \propto \int_{\psi(r,\theta=0^+)=\psi'(r)}^{\psi(r,\theta=2\pi^-)=\psi(r)} [d\psi] \psi(re^{i\theta}) e^{-S_E^R} \propto \langle\psi|e^{-2\pi K_{\text{pen}}} \psi(re^{i\theta})|\psi'\rangle$. To obtain the most general form, $\psi$ needs to smeared out on the Euclidean plane. $S_E^R$ is the Euclidean action for Rindler coordinates [4].

From now, we will write $K := K_{\text{tot}}$ and

$$\sigma(\lambda) = A^{\frac{1}{2}} \rho^{(0)} O(\lambda), \quad (12)$$

and thus,

$$O(\lambda) = \sum_{\mu,\nu} \left[ \int dr d\theta f_{\mu\nu}(r,\theta) \psi(re^{i\theta} - \lambda) \otimes E_{\mu\nu}(\theta) \right]. \quad (13)$$

Note: We can substitute $O(\lambda)$ for $\tilde{O}_p$ to define $\rho(\lambda)$ in Eq. (6).

## IV. EVALUATION OF $S^{(2)''}$

Our state [Eq. (9)] is in the form,

$$\rho(\lambda) = \rho_{\text{pen}}^{(0)} \otimes \rho_{\text{aux}}^{(0)} + \sigma(\lambda). \quad (14)$$

We can expand the entanglement entropy perturbatively in $\sigma$,

$$S_{\text{out}}(\lambda) = S^{(0)}(\lambda) + S^{(1)}(\lambda) + S^{(2)}(\lambda) + \ldots, \quad (15)$$

where $S^{(n)}$ contains $n$ powers of $\sigma$. We assume without loss of generality that $\rho_{\text{pen}}^{(0)} \otimes \rho_{\text{aux}}^{(0)}$ is normalized and $\text{Tr}(\sigma) = 0$ so that $\rho(\lambda)$ is also normalized.

It has been shown that under these conditions [11,16],

$$(S^{(0)} + S^{(1)})'' = \frac{2\pi A}{\hbar} \langle T_{kk} \rangle, \quad (16)$$

where we define $f'' = \frac{d^2}{d\lambda^2} f(\lambda)|_{\lambda=0}$ for any function $f$.

Subtracting $S''_{\text{out}}$ from both sides of Eq. (16) and slightly rearranging gives [11,16]

$$\frac{\hbar}{2\pi A} S''_{\text{out}} - \langle T_{kk} \rangle = \frac{\hbar}{2\pi A} (S_{\text{out}} - S^{(0)} - S^{(1)})'' \\ = \frac{\hbar}{2\pi A} S^{(2)''} + \ldots, \quad (17)$$

where terms in "…" contain terms higher than the quadratic order in $\sigma$ or equivalently, higher than the zeroth order in $A$.

The QNEC states that the left-hand side of Eq. (17) is negative in the limit as $A \to 0$. In this limit, only the first term ($\frac{\hbar}{2\pi A} S^{(2)''}$) on the right-hand side is nonzero, and so to prove QNEC, it is sufficient to prove that $S^{(2)''} \leq 0$ for perturbations about the vacuum.

Note: In general, $\sigma$ will contain terms proportional to $A^{\frac{n}{2}}$ for $n \geq 1$. However, only the terms proportional to $A^{\frac{1}{2}}$ are relevant for proving the QNEC as contributions of higher order terms will drop out as $A \to 0$ in Eq. (17). Hence, we can ignore the "…" terms in Eq. (5).

As in the bosonic case [16], we will use the replica trick to calculate $S^{(2)''}$. The replica trick is used to calculate entanglement entropies as [21]

$$S_{\text{out}} = -\text{Tr}[\rho \log \rho] = (1 - n\partial_n) \log \text{Tr}[\rho^n]|_{n=1} \quad (18)$$

$$= D \log \tilde{Z}_n, \quad (19)$$

where $\tilde{Z}_n = \text{Tr}[\rho^n]$ and $D$ is an operator defined by

$$Df(n) := (1 - n\partial_n)f(n)|_{n=1} \quad (20)$$

for some function $f(n)$. Notice that to apply Eq. (19), we must analytically continue $\tilde{Z}_n$ to real $n > 0$.

It has been shown [16] that at the quadratic order in $\sigma$, $\log \tilde{Z}_n$ can be written as

$$\log \tilde{Z}_n \supset -\frac{n}{2}\langle OO \rangle_n + \frac{1}{2}\left\langle \left(\sum_{k=0}^{n-1} O^{(k)}\right)^2 \right\rangle_n, \quad (21)$$

where

$$\langle \ldots \rangle_n := \frac{\text{Tr}[(\rho^{(0)})^n T[\ldots]]}{\text{Tr}[(\rho^{(0)})^n]}. \quad (22)$$

$T[\ldots]$ is the $\theta$ ordering, and

$$O^{(k)} := (\rho^{(0)})^{-k} O (\rho^{(0)})^k \quad (23)$$

$$= e^{2\pi k K} O e^{-2\pi k K}. \quad (24)$$

This is equivalent to the Heisenberg evolution of $O$ in the angle $\theta$ by $2\pi k$. We can obtain $O^{(k)}$ from $O$ by letting the range of integration that defines $O$ shift from $[0, 2\pi]$ to $[2\pi k, 2\pi(k+1)]$ as long as we define $f_{\mu\nu}(r,\theta)$ to be antiperiodic with period $2\pi^6$ [16]. Hence,

$$S^{(2)\prime\prime} = D\frac{-n}{2}\langle OO \rangle_n'' + D\frac{1}{2}\left\langle \left(\sum_{k=0}^{n-1} O^{(k)}\right)^2 \right\rangle_n''. \quad (25)$$

### A. The correlators

The same-sheet correlator, $D\frac{-n}{2}\langle OO \rangle_n''$, has previously been evaluated [16] to give

$$D\frac{-n}{2}\langle OO \rangle_n'' = \frac{1}{2}\langle OOT(0) \rangle, \quad (26)$$

where $T(x) = -2\pi A T_{kk}(x)$, $T(x)$ being the energy-momentum tensor of the CFT [17], and we write $\langle \cdots \rangle := \langle \cdots \rangle_1$. Explicitly, using Eq. (13), this can be written as

$$\frac{1}{2}\langle OOT(0) \rangle = \frac{1}{2(2\pi)^2} \sum_{\substack{\mu,\nu,\mu',\nu' \\ m,m'}} \int dr dr' d\theta d\theta' [f_{\mu\nu}^{(m)}(r) f_{\mu'\nu'}^{(m')}(r')$$

$$\times e^{-im(\theta+\frac{1}{2})} e^{-im'(\theta'+\frac{1}{2})} \langle \psi(re^{i\theta}) \psi(r'e^{i\theta'}) T(0) \rangle$$

$$\times \langle E_{\mu\nu}(\theta) E_{\mu'\nu'}(\theta') \rangle], \quad (27)$$

where we Fourier expand $f_{\mu\nu}(r,\theta)$ as

$$f_{\mu\nu}(r,\theta) = \frac{1}{2\pi} \sum_{m=-\infty}^{m=\infty} f_{\mu\nu}^{(m)}(r) e^{-i(m+\frac{1}{2})\theta}. \quad (28)$$

Notice that the extra $\frac{1}{2}$ term in the exponential automatically makes $f_{\mu\nu}(r,\theta)$ antiperiodic as required. In Fourier components, the reality on $f_{\mu\nu}(r,\theta)$ becomes

$$f_{\mu\nu}^{(m)}(r) = i f_{\nu\mu}^{*(m)}(r). \quad (29)$$

Using Eqs. (A9) and (A25) from the Appendix for the two-point correlation functions, Eq. (27) gives

$$\frac{1}{2}\langle OOT(0) \rangle = -\frac{1}{2(2\pi)^3} \sum_{\mu,\nu,m,m',p} \int \frac{drdr'd\theta d\theta'}{(rr')^2}\left\{ \right. \quad (30)$$

$$[rf_{\mu\nu}^{(m)}(r) f_{\nu\mu}^{(m')}(r') e^{i\theta(-p-m-2)} e^{-i\theta'(-p-m'-2)}$$

$$- r' f_{\mu\nu}^{(m)}(r) f_{\nu\mu}^{(m')}(r') e^{i\theta(-p-m-3)} e^{-i\theta'(-p-m'-1)}]$$

$$\times e^{\pi(K_\nu + K_\mu)} \frac{\cosh(\pi \alpha_{\mu\nu})}{i(p+\frac{1}{2}) + \alpha_{\mu\nu}}\right\}, \quad (31)$$

where we define $\alpha_{\mu\nu} := K_\nu - K_\mu$. For $n = 1$, $p \in \mathbb{Z}$ so that after doing the angle integration, using the Kronecker delta from the integration, redefining the dummy variable $m \to m - 2$ for the first term and $m \to m - 3$ for the second term, we find that

$$\frac{1}{2}\langle OOT(0) \rangle$$

$$= \frac{1}{4\pi} \sum_{\mu,\nu,m} \int \frac{drdr'}{(rr')^2} [r f_{\mu\nu}^{(m-2)}(r) f_{\nu\mu}^{(-m-2)}(r')$$

$$- r' f_{\mu\nu}^{(m-3)}(r) f_{\nu\mu}^{(-m-1)}(r')] e^{\pi(K_\nu + K_\mu)} \frac{\cosh(\pi \alpha_{\mu\nu})}{i(m-\frac{1}{2}) - \alpha_{\mu\nu}}. \quad (32)$$

This can be further simplifed by redefining the dummy variables for the second term with

$$r \leftrightarrow r' \quad \mu \leftrightarrow \nu \quad m \to -m + 1, \quad (33)$$

giving

---

[6]After a shift by $2\pi$, $\psi$ picks up an extra minus sign [see Eq. (A8)]. To cancel this negative sign, $f_{\mu\nu}(r,\theta)$ needs to be defined antiperiodically [i.e., $f_{\mu\nu}(r,\theta + 2\pi k) = (-1)^k f_{\mu\nu}(r,\theta)$].

$$D\frac{-n}{2}\langle OO\rangle''_n = \frac{1}{2\pi}\sum_{\mu,\nu,m}\int\frac{drdr'}{(rr')^2}rf^{(m-2)}_{\mu\nu}(r)f^{(-m-2)}_{\nu\mu}(r')$$

$$\times e^{\pi(K_\nu+K_\mu)}\frac{\cosh(\pi\alpha_{\mu\nu})}{i(m-\frac{1}{2})-\alpha_{\mu\nu}}. \quad (34)$$

Focusing on the multisheet correlator from Eq. (25), $D\frac{1}{2}\langle(\sum_{k=0}^{n-1}O^{(k)})^2\rangle''_n$, we have that

$$\left\langle\left(\sum_{k=0}^{n-1}O^{(k)}\right)^2\right\rangle_n$$
$$=\left\langle\left[\sum_{\mu\nu}\int_0^{2\pi n}f_{\mu\nu}(r,\theta)\psi(r,\theta)\otimes E_{\mu\nu}(\theta)\right]^2\right\rangle_n. \quad (35)$$

The traces in Eq. (22) can be evaluated via a path integral on a $n$-replicated manifold, and so we interpret $O^{(k)}$ as $O$ inserted on the $(k+1)$th replica sheet. The $n$-replicated manifold is a Riemann manifold constructed by $n$ copies of the Euclidean plane with various line segments identified (see Fig. 3). Thus, the sums over $n$ and integrations over $\theta \in [0,2\pi]$ can be replaced with one integration over $\theta \in [0,2\pi n]$ [see Eq. (35)], as this covers the whole replicated manifold. $\psi = \psi(r,\theta)$ is now a chiral left-moving fermionic field defined on the replicated manifold rather than the Euclidean plane. The definition of $\psi$ for any angle is still given by the Heisenberg evolution rule from Eq. (A8) [16]. $f(r,\theta)$ is still antiperiodic and defined via its Fourier expansion for all angles.

Due to the different boundary conditions depending on whether $n$ is even or odd, the correlation functions have to be evaluated separately for each case [19,22]. Instead, we will only focus on the case that $n$ is odd and use this to analytically continue our expression to the positive real line. For now, we will assume that $n$ is odd.

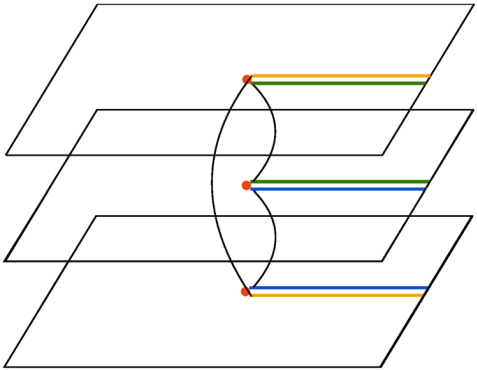

FIG. 3. An $n$-replicated manifold for $n=3$ constructed from three copies of the Euclidean plane with various line segments identified with color coding.

Hence, using the above, we have

$$D\frac{1}{2}\left\langle\left(\sum_{k=0}^{n-1}O^{(k)}\right)^2\right\rangle''_n$$
$$=D\frac{1}{2(2\pi)^2}\sum_{\mu,\nu,m,m'}\int drdr'd\theta d\theta'$$
$$\times[f^{(m)}_{\mu\nu}(r)f^{(m')}_{\mu\nu}(r')e^{-i(m-\frac{1}{2})\theta}e^{-i(m'-\frac{1}{2})\theta'}$$
$$\times\langle\psi(r,\theta)\psi(r',\theta')\rangle''_n\langle E_{\mu\nu}(\theta)E_{\nu\mu}(\theta')\rangle_n]. \quad (36)$$

We show in the Appendix that,

$$\langle\psi(z)\psi(w)\rangle''_n$$
$$=\frac{-1}{4n(zw)^{\frac{3}{2}}}\sum_{|q|\leq 1}\text{sign}(q)(4q^2-1)\left(\frac{w}{z}\right)^q \quad (37)$$

$$=\frac{-1}{4n(rr')^{\frac{3}{2}}}\sum_{|q|\leq 1}\text{sign}(q)P\left(q,\frac{r'}{r}\right)e^{i\theta(-q-\frac{3}{2})}e^{i\theta'(q-\frac{3}{2})}, \quad (38)$$

where $q \in \frac{\mathbb{Z}}{n}+\frac{1}{2n}$ (i.e., $q$ is summed over some odd integers divided by $2n$) and

$$P(q,R)=(4q^2-1)R^q. \quad (39)$$

Notice that when $n=1$, $q$ is summed over just two values ($\frac{1}{2}$ and $-\frac{1}{2}$) for which $P=0$. This is expected since for $n=1$, we have translation invariance, and so the CFT correlator should not depend on $\lambda$. This fact will be useful later.

Also, we show in the Appendix that

$$\langle E_{\mu\nu}(\theta)E_{\mu'\nu'}(\theta')\rangle_n$$
$$=-\frac{1}{\pi nZ^{\bar{a}ux}_n}\sum_p\left[e^{-i(\theta-\theta')(p+\frac{1}{2})}\right.$$
$$\left.\times\frac{\cosh(n\pi\alpha_{\mu\nu})}{i(p+\frac{1}{2})+\alpha_{\mu\nu}}e^{-\pi n(K_\nu+K_\mu)}\delta_{\mu\nu'}\delta_{\mu'\nu}\right], \quad (40)$$

where $p \in \frac{\mathbb{Z}}{n}$.

Putting everything together into Eq. (36), we get

$$D\frac{1}{4n^2(2\pi)^3}\sum_{\substack{\nu,\mu,m,m'\\p,|q|<1}}\int_0^{2\pi n}\int_0^{2\pi n}\frac{drdr'd\theta d\theta'}{(rr')^{\frac{3}{2}}}$$
$$\times\left[f^{(m)}_{\mu\nu}(r)f^{(m')}_{\nu\mu}(r')e^{i\theta(-p-q-m-\frac{5}{2})}e^{-i\theta'(p+q-m'-\frac{3}{2})}\right.$$
$$\left.\times\frac{\text{sign}(q)P(q,\frac{r'}{r})}{i(p+\frac{1}{2})+\alpha_{\mu\nu}}\times e^{-n\pi(K_\mu+K_\nu)}\cosh(n\pi\alpha_{\mu\nu})\right]. \quad (41)$$

Notice that $(-p-q-m-\frac{5}{2})$ and $(p+q-m'-\frac{3}{2}) \in \frac{\mathbb{Z}}{n}$. To see this, we only need to show $-q-\frac{5}{2}$ and $q-\frac{3}{2} \in \frac{\mathbb{Z}}{n}$ since $-p-m$ and $p-m'$ are already in $\frac{\mathbb{Z}}{n}$. Focusing on $-q-\frac{5}{2}$, we rewrite $q = \frac{I}{n} + \frac{1}{2n}$ where $I \in \mathbb{Z}$. Thus,

$$-\frac{I}{n} - \frac{1}{2n} - \frac{5}{2} = \frac{-2I + 5n - 1}{2n}. \tag{42}$$

Since $n \geq 1$ and odd, we can write $n = 2t+1$ for integer $t \geq 0$ so that

$$\frac{-2I+5n-1}{2n} = \frac{-2I+5(2t+1)-1}{2n}$$
$$= \frac{-I+5t+2}{n} \in \frac{\mathbb{Z}}{n}. \tag{43}$$

A similar result holds for $q - \frac{3}{2}$.

This means that doing the angle integration in Eq. (41) gives a Kronecker delta multiplied by $2\pi n$. After relabeling $m \to m - 2$, we have

$$D\frac{1}{2}\left\langle \left(\sum_{k=0}^{n-1} O^{(k)}\right)^2 \right\rangle''_n$$
$$= \frac{i}{(8\pi)} \sum_{\mu,\nu,m} \int \frac{drdr'}{(rr')^{\frac{3}{2}}} \left\{ f_{\mu\nu}^{(m-2)}(r) f_{\nu\mu}^{(-m-2)}(r') \right.$$
$$\left. \times D\left[\sum_{|q|<1} \frac{\text{sign}(q)P(q,\frac{r'}{r})}{(q+m)+i\alpha_{\mu\nu}}\right] e^{-\pi(K_\mu+K_\nu)} \cosh(\pi\alpha_{\nu\mu}) \right\}. \tag{44}$$

We have moved the operator $D$ inside the integral and set $n = 1$ for terms outside $D$. To do this, we used the fact that for any two functions $f(n)$ and $g(n)$ such that $f(1)$ and $\frac{d}{dn}f(n)|_{n=1}$ are finite and $g(n) = 0$, we have [16]

$$D(f(n)g(n)) = f(1)Dg(n). \tag{45}$$

In Eq. (44), $g(n)$ is the sum over $q$ inside $D$, which we already showed above is zero for $n = 1$. To apply the operator $D$, we must analytically continue $g(n)$. The method we use to analytically continue $g(n)$ is similar to the method used for the bosonic case [16].

### B. Evaluation of operator $D$

To analytically continue the expression inside $D$, notice that

$$\sum_{|q|<1} \left[\frac{\text{sign}(q)P(q,\frac{r'}{r})}{(q+m+i\alpha_{\mu\nu})}\right]$$
$$= \sum_{\frac{1}{2n}\leq q \leq 1-\frac{1}{2n}} \left[\frac{P(q,\frac{r'}{r})}{(q+m+i\alpha_{\mu\nu})} + \frac{P(-q,\frac{r'}{r})}{(q-m-i\alpha_{\mu\nu})}\right]$$
$$= \sum_{\frac{1}{2n}\leq q \leq \frac{1}{2}-\frac{1}{n}} \left[\frac{P(q,\frac{r'}{r})}{q-z_m} + \frac{P(-q,\frac{r'}{r})}{q+z_m}\right]$$
$$+ \sum_{\frac{1}{2}+\frac{1}{n}\leq q \leq 1-\frac{1}{2n}} \left[\frac{P(q,\frac{r'}{r})}{q-z_m} + \frac{P(-q,\frac{r'}{r})}{q+z_m}\right], \tag{46}$$

where in the first line, we removed the sign$(q)$ term by rewriting the sum over $0 < q < 1$ or equivalently over $\frac{1}{2n} \leq q < 1 - \frac{1}{2n}$. In the second line, we split the sum over $q$ into two sum over $\frac{1}{2n} < q < \frac{1}{2} - \frac{1}{n}$ and $\frac{1}{2} + \frac{1}{n} < q < 1 - \frac{1}{n}$. Recall that $P(\pm\frac{1}{2}, \frac{r'}{r}) = 0$, and so we can remove the sum over $q = \frac{1}{2}$. We have also defined $z_m := -m - i\alpha_{\mu\nu}$.

By writing $q = \frac{k}{n}$, we can rewrite the sum in Eq. (46) as

$$\sum_{|q|<1}\left[\frac{\text{sign}(q)P(q,\frac{r'}{r})}{(q+m+i\alpha_{\mu\nu})}\right]$$
$$= \sum_{\frac{1}{2}\leq k \leq \frac{n}{2}-1}\left[\frac{P(\frac{k}{n},\frac{r'}{r})}{\frac{k}{n}-z_m} + \frac{P(-\frac{k}{n},\frac{r'}{r})}{\frac{k}{n}+z_m}\right]$$
$$+ \sum_{\frac{n}{2}+1\leq k \leq n-\frac{1}{2}}\left[\frac{P(\frac{k}{n},\frac{r'}{r})}{\frac{k}{n}-z_m} + \frac{P(-\frac{k}{n},\frac{r'}{r})}{\frac{k}{n}+z_m}\right], \tag{47}$$

where $k \in \mathbb{Z} + \frac{1}{2}$.

#### 1. First sum

Focusing on the first sum of Eq. (47), we need to find

$$D \sum_{\frac{1}{2}\leq k \leq \frac{n}{2}-1}\left[\frac{P(\frac{k}{n},\frac{r'}{r})}{\frac{k}{n}-z_m} + \frac{P(-\frac{k}{n},\frac{r'}{r})}{\frac{k}{n}+z_m}\right]. \tag{48}$$

We can extend $P(z) := P(z,\frac{r'}{r})$ to complex $z$, making $P(z)$ an analytical function, independent of $n$. We can write Eq. (48) as [16]

$$D \sum_{\frac{1}{2}\leq k \leq \frac{n}{2}-1}\left[\frac{P(\frac{k}{n},\frac{r'}{r}) - P(z_m)}{\frac{k}{n}-z_m} + \frac{P(-\frac{k}{n},\frac{r'}{r}) - P(z_m)}{\frac{k}{n}+z_m}\right]$$
$$+ D \sum_{\frac{1}{2}\leq k \leq \frac{n}{2}-1}\left[\frac{P(z_m)}{\frac{k}{n}-z_m} + \frac{P(z_m)}{\frac{k}{n}+z_m}\right] \tag{49}$$

and evaluate the two terms separately.

We can write

$$D \sum_{\frac{1}{2}\leq k\leq \frac{n}{2}-1}\left[\frac{P(\frac{k}{n}) - P(z_m)}{\frac{k}{n} - z_m}\right]$$
$$= D \sum_{\frac{1}{2}\leq k \leq \frac{n}{2}-1} \sum_{r=1}^\infty b_r \frac{(\frac{k}{n}+\frac{1}{2})^r - (z_m+\frac{1}{2})^r}{(\frac{k}{n}+\frac{1}{2}) - (z_m+\frac{1}{2})}, \tag{50}$$

where we obtain the second line by expanding $P(z)$ about $z = -\frac{1}{2}$ as

$$P(z) = \sum_{r=0}^\infty b_r \left(z + \frac{1}{2}\right)^r. \tag{51}$$

After using the following identity:

$$\frac{x^r - z^r}{x - z} = \sum_{s=0}^{r-1} z^{r-s-1} x^s, \quad (52)$$

we can write Eq. (50) as

$$\sum_{r=1}^{\infty} \sum_{s=0}^{r-1} b_r \left(z_m + \frac{1}{2}\right)^{r-s-1} D \sum_{\frac{1}{2} \leq k \leq \frac{n}{2}-1} \left(\frac{k}{n} + \frac{1}{2}\right)^s. \quad (53)$$

To evaluate $D \sum_{\frac{1}{2} \leq k \leq \frac{n}{2}-1} \left(\frac{k}{n} + \frac{1}{2}\right)^s$, we write $n$ as $n = 2t + 1$ for integer $t$ and relabel $k \to k - \frac{1}{2}$ so that

$$D \sum_{\frac{1}{2} \leq k \leq \frac{n}{2}-1} \left(\frac{k}{n} + \frac{1}{2}\right)^s = D \sum_{k=1}^{t} \left(\frac{k - \frac{1}{2}}{2t + 1} + \frac{1}{2}\right)^s, \quad (54)$$

where now $k \in \mathbb{Z}$ in the second term. The expression inside $D$ can now be analytically continued. Using the Hurwitz zeta function $\zeta(s, a)$, where

$$\zeta(s, a) = \sum_{k=0}^{\infty} \frac{1}{(k + a)^s}, \quad (55)$$

we can write Eq. (54) as

$$D \left[\frac{1}{(2t+1)}\right]^s [\zeta(-s, t+1) - \zeta(-s, 2t+1)]$$
$$= D[\zeta(-s, t+1) - \zeta(-s, 2t+1)], \quad (56)$$

where to get to the second line, we applied Eq. (45). The Hurwitz zeta function is analytic in $a$ which means that we can apply the operator $D$. Recall that $D = (1 - n\partial_n)|_{n=1} = (1 - \frac{(2t+1)}{2}\partial_t)|_{t=0}$ and that the expression inside $D$ is zero when $t = 0$, so we can simplify Eq. (56) as

$$-\frac{1}{2} \partial_t [\zeta(-s, t+1) - \zeta(-s, 2t+1)]|_{t=0}. \quad (57)$$

Using $\partial_a \zeta(s, a) = -s\zeta(s+1, a)$, we find that Eq. (56) can be evaluated as

$$-\frac{1}{2} [s\zeta(-s+1, 1) - 2s\zeta(-s+1, 1)]$$
$$= \frac{1}{2} s\zeta(-s+1, 1) = \frac{1}{2} s\zeta(1-s), \quad (58)$$

where $\zeta(s) = \zeta(s, 1)$ is the Riemann zeta function.

Putting everything together, we find that

$$D \sum_{\frac{1}{2} \leq k \leq \frac{n}{2}-1} \left[\frac{P(\frac{k}{n}) - P(z_m)}{\frac{k}{n} - z_m}\right]$$
$$= \frac{1}{2} \sum_{r=1}^{\infty} \sum_{s=0}^{r-1} b_r \left(z_m + \frac{1}{2}\right)^{r-s-1} s\zeta(1-s). \quad (59)$$

Similarly, for the $\frac{P(-\frac{k}{n}) - P(z_m)}{\frac{k}{n} + z_m}$ term from Eq. (49), we can write

$$D \sum_{\frac{1}{2} \leq k \leq \frac{n}{2}-1} \left[\frac{P(-\frac{k}{n}) - P(z_m)}{\frac{k}{n} + z_m}\right]$$
$$= -D \sum_{\frac{1}{2} \leq k \leq \frac{n}{2}-1} \sum_{r=1}^{\infty} a_r \frac{(-\frac{k}{n} - \frac{1}{2})^r - (z_m - \frac{1}{2})^r}{(-\frac{k}{n} - \frac{1}{2}) - (z_m - \frac{1}{2})}, \quad (60)$$

where we expanded $P(z) = \sum_{r=0}^{\infty} a_r (z - \frac{1}{2})^r$. The above can be written as

$$-\sum_{r=1}^{\infty} \sum_{s=0}^{r-1} (-1)^s a_r \left(z_m - \frac{1}{2}\right)^{r-s-1} D \sum_{\frac{1}{2} \leq k \leq \frac{n}{2}-1} \left(\frac{k}{n} + \frac{1}{2}\right)^s \quad (61)$$

and can be evaluated to give

$$-\frac{1}{2} \sum_{r=1}^{\infty} \sum_{s=0}^{r-1} a_r \left(z_m - \frac{1}{2}\right)^{r-s-1} s\zeta(1-s)(-1)^s. \quad (62)$$

Now, we need to evaluate the last term of the first sum; i.e., we need to evaluate from Eq. (49),

$$D \sum_{\frac{1}{2} \leq k \leq \frac{n}{2}-1} \left(\frac{P(z_m)}{\frac{k}{n} - z_m} + \frac{P(z_m)}{\frac{k}{n} + z_m}\right). \quad (63)$$

We can write this as [16]

$$P(z_m) D \sum_{k=1}^{t} \left(\frac{1}{k - \frac{1}{2} - z_m(2t+1)} + \frac{1}{k - \frac{1}{2} + z_m(2t+1)}\right), \quad (64)$$

where we have already relabeled $k$ so that $k \in \mathbb{Z}$. This expression can be written in terms of digamma functions, $\psi(z)$, as

$$P(z_m) D \left\{ \psi\left[t + \frac{1}{2} - z_m(2t+1)\right] - \psi\left[\frac{1}{2} - z_m(2t+1)\right] \right.$$
$$\left. + \psi\left[t + \frac{1}{2} + z_m(2t+1)\right] - \psi\left[\frac{1}{2} + z_m(2t+1)\right] \right\}, \quad (65)$$

where in terms of the Gamma function $\Gamma(z)$,

$$\psi(z) = \frac{\Gamma'(z)}{\Gamma(z)} = -\gamma + \sum_{k=0}^{\infty} \left( \frac{1}{k+1} - \frac{1}{k+z} \right). \quad (66)$$

However, we can not apply $D$ straight away since we need to select the correct analytic continuation to real positive $t$. The digamma function has poles at zero and all negative integers, and so the expression in Eq. (64) may have poles for real positive $t$.

The method used to select the correct analytic continuation has already been done for the bosonic case [16], which avoids poles along real positive $t$ and can straightforwardly be applied to our case with some minor modifications. One has to consider the three cases $m < 0$, $m = 0$ and $m > 0$ separately, which leads to

$$D \sum_{\frac{1}{2} \leq k \leq \frac{n}{2}-1} \left( \frac{P(z_m)}{\frac{k}{n} - z_m} + \frac{P(z_m)}{\frac{k}{n} + z_m} \right)$$
$$= -\frac{1}{2} P(z_m) \delta(m) \pi^2 \text{sech}^2(\pi \alpha_{\mu\nu}). \quad (67)$$

Putting everything together from Eq. (59), (62), and (67), we can write the first sum [Eq. (48)] as

$$\frac{1}{2} \sum_{r=1}^{\infty} \sum_{s=0}^{r-1} b_r \left( z_m + \frac{1}{2} \right)^{r-s-1} s\zeta(1-s)$$
$$- \frac{1}{2} \sum_{r=1}^{\infty} \sum_{s=0}^{r-1} a_r \left( z_m - \frac{1}{2} \right)^{r-s-1} s\zeta(1-s)(-1)^s$$
$$- \frac{1}{2} P(z_m) \delta(m) \pi^2 \text{sech}^2(\pi \alpha_{\mu\nu}). \quad (68)$$

### 2. Second sum

Finally, focusing on the second sum of Eq. (47), we find that using similar methods for the first sum,

$$D \sum_{\frac{n}{2}+1 \leq k \leq n-\frac{1}{2}} \left[ \frac{P(\frac{k}{n}) - P(z_m)}{\frac{k}{n} - z_m} \right]$$
$$= \sum_{r=1}^{\infty} \sum_{s=0}^{r-1} a_r \left( z_m - \frac{1}{2} \right)^{r-s-1} D \sum_{\frac{n}{2}+1 \leq k \leq n-\frac{1}{2}} \left( \frac{k}{n} - \frac{1}{2} \right)^s$$
$$= \frac{1}{2} \sum_{r=1}^{\infty} \sum_{s=0}^{r-1} a_r \left( z_m - \frac{1}{2} \right)^{r-s-1} s\zeta(1-s) \quad (69)$$

and

$$D \sum_{\frac{n}{2}+1 \leq k \leq n-\frac{1}{2}} \left[ \frac{P(-\frac{k}{n}) - P(z_m)}{\frac{k}{n} + z_m} \right]$$
$$= -\frac{1}{2} \sum_{r=1}^{\infty} \sum_{s=0}^{r-1} b_r \left( z_m + \frac{1}{2} \right)^{r-s-1} s\zeta(1-s)(-1)^s. \quad (70)$$

And after picking out the correct analytic continuation,

$$D \sum_{\frac{n}{2}+1 \leq k \leq n-\frac{1}{2}} \left[ \frac{P(z_m)}{\frac{k}{n} - z_m} + \frac{P(z_m)}{\frac{k}{n} + z_m} \right]$$
$$= \frac{1}{2} P(z_m) \left[ \frac{1}{(z-\frac{1}{2})^2} + \frac{1}{(z+\frac{1}{2})^2} - \delta(m)\pi^2 \text{sech}^2(\pi \alpha_{\mu\nu}) \right]. \quad (71)$$

Putting all the terms from the first and second sum together into Eq. (47) gives

$$D \left[ \sum_{|q|<1} \frac{\text{sign}(q) P(q, \frac{r'}{r})}{(q+m) + i\alpha_{\mu\nu}} \right] \quad (72a)$$

$$= \frac{1}{2} \sum_{r=1}^{\infty} \sum_{s=0}^{r-1} a_r \left( z_m - \frac{1}{2} \right)^{r-s-1} (1-(-1)^s) s\zeta(1-s) \quad (72b)$$

$$+ \frac{1}{2} \sum_{r=1}^{\infty} \sum_{s=0}^{r-1} b_r \left( z_m + \frac{1}{2} \right)^{r-s-1} (1-(-1)^s) s\zeta(1-s) \quad (72c)$$

$$+ \frac{1}{2} P(z_m) \left[ \frac{1}{(z_m - 1/2)^2} + \frac{1}{(z_m + 1/2)^2} \right]$$
$$- P(z_m) \delta(m) \pi^2 \text{sech}^2(\pi \alpha_{\mu\nu}). \quad (72d)$$

Notice that for $s$ even, $(1-(-1)^s) = 0$ and for $s > 1$ and odd, $\zeta(1-s) = 0$. Hence, only when $s = 1$ does the sum above over $s$ contribute in Eqs. (72b) and (72c). We can rewrite these sums over $s$ as [16]

$$\zeta(0) \sum_{r=2}^{\infty} a_r \left( z_m - \frac{1}{2} \right)^{r-2}$$
$$= -\frac{1}{2} \left[ \frac{P(z_m)}{(z_m - \frac{1}{2})^2} - \frac{a_1}{(z_m - \frac{1}{2})} - \frac{a_0}{(z_m - \frac{1}{2})^2} \right] \quad (73)$$

and

$$\zeta(0) \sum_{r=2}^{\infty} b_r \left( z_m + \frac{1}{2} \right)^{r-2}$$
$$= -\frac{1}{2} \left[ \frac{P(z_m)}{(z_m + \frac{1}{2})^2} - \frac{b_1}{(z_m + \frac{1}{2})} - \frac{b_0}{(z_m + \frac{1}{2})^2} \right]. \quad (74)$$

Recall that $P = P(z_m, \frac{r'}{r})$ given by Eq. (39). Hence, $a_0 = b_0 = 0$, $a_1 = 4(\frac{r'}{r})^{\frac{1}{2}}$, and $b_1 = -4(\frac{r'}{r})^{-\frac{1}{2}}$. This means that Eq. (72a) can be further simplified to give

$$D \left[ \sum_{|q|<1} \frac{\text{sign}(q) P(q, \frac{r'}{r})}{(q+m) + i\alpha_{\mu\nu}} \right]$$
$$= \frac{2(\frac{r'}{r})^{\frac{1}{2}}}{(z_m - \frac{1}{2})} - \frac{2(\frac{r'}{r})^{-\frac{1}{2}}}{(z_m + \frac{1}{2})} - P(z_m) \delta(m) \pi^2 \text{sech}^2(\pi \alpha_{\mu\nu}). \quad (75)$$

### C. Showing $S^{(2)''} \leq 0$

Plugging Eq. (75) back into Eq. (44), we get for the multisheet correlator

$$D\frac{1}{2}\left\langle\left(\sum_{k=0}^{n-1} O^{(k)}\right)^2\right\rangle_n''$$
$$= \frac{i}{(4\pi)}\sum_{\mu,\nu,m}\int \frac{drdr'}{(rr')^{\frac{3}{2}}}\left\{f^{(m-2)}_{\mu\nu}(r)f^{(-m-2)}_{\nu\mu}(r')\right.$$
$$\times\left[\frac{(\frac{r'}{r})^{\frac{1}{2}}}{(-m-i\alpha_{\mu\nu}-\frac{1}{2})} - \frac{(\frac{r'}{r})^{-\frac{1}{2}}}{(-m-i\alpha_{\mu\nu}+\frac{1}{2})}\right]$$
$$\left.\times e^{-\pi(K_\mu+K_\nu)}\cosh(\pi\alpha_{\nu\mu})\right\} \quad (76a)$$

$$+\frac{i}{(8\pi)}\sum_{\mu,\nu,m}\int \frac{drdr'}{(rr')^{\frac{3}{2}}}[f^{(m-2)}_{\mu\nu}(r)f^{(-m-2)}_{\nu\mu}(r')$$
$$\times -P(-m-i\alpha_{\mu\nu})\delta(m)\pi^2\mathrm{sech}^2(\pi\alpha_{\mu\nu})$$
$$\times e^{-\pi(K_\mu+K_\nu)}\cosh(\pi\alpha_{\nu\mu})]. \quad (76b)$$

After expanding and relabeling ($r \leftrightarrow r'$, $\mu\nu \leftrightarrow \nu\mu$, and $m \to -m$) for the first sum in Eq. (76a), we find that Eq. (76a) simplifies to

$$-\frac{1}{(2\pi)}\sum_{\mu,\nu,m}\int \frac{drdr'}{(rr')^2}\left[rf^{(m-2)}_{\mu\nu}(r)f^{(-m-2)}_{\nu\mu}(r')\right.$$
$$\left.\times\frac{1}{i(m-\frac{1}{2})-\alpha_{\mu\nu}}e^{-\pi(K_\mu+K_\nu)}\cosh(\pi\alpha_{\nu\mu})\right]. \quad (77)$$

But this term cancels with Eq. (34) from the single sheet correlator. Hence, we find that

$$S^{(2)''} = \frac{i}{(8\pi)}\sum_{\mu,\nu}\int \frac{drdr'}{(rr')^{\frac{3}{2}}}[f^{(-2)}_{\mu\nu}(r)f^{(-2)}_{\nu\mu}(r')$$
$$\times -P(-i\alpha_{\mu\nu})\pi^2\mathrm{sech}(\pi\alpha_{\mu\nu})\times e^{-\pi(K_\mu+K_\nu)}]. \quad (78)$$

Using $P(-i\alpha_{\mu\nu}) = -(4\alpha^2_{\mu\nu}+1)(\frac{r'}{r})^{-i\alpha_{\mu\nu}}$ and from the reality condition, $f^{(-2)}_{\nu\mu}(r) = if^{*(-2)}_{\mu\nu}(r)$, we have that

$$S^{(2)''} = -\frac{1}{(8\pi)}\sum_{\mu,\nu}\int \frac{drdr'}{(rr')^{\frac{3}{2}}}\left[f^{(-2)}_{\mu\nu}(r)f^{*(-2)}_{\mu\nu}(r')\right.$$
$$\left.\times(4\alpha^2_{\mu\nu}+1)\left(\frac{r'}{r}\right)^{-i\alpha_{\mu\nu}}\pi^2\mathrm{sech}(\pi\alpha_{\mu\nu})\times e^{-\pi(K_\mu+K_\nu)}\right]$$
$$= -\frac{\pi}{8}\sum_{\mu,\nu}\left[\int drr^{-\frac{3}{2}+i\alpha_{\mu\nu}}f^{(-2)}_{\mu\nu}(r)\right]^2$$
$$\times(4\alpha^2_{\mu\nu}+1)\mathrm{sech}(\pi\alpha_{\mu\nu})e^{-\pi(K_\mu+K_\nu)} \leq 0. \quad (79)$$

Thus, $S^{(2)''} \leq 0$ which proves the QNEC.

## V. DISCUSSION

It has been shown that the above proof can be extended to $D=2$ by dimensional reduction [16]. Therefore, QNEC also holds for a free fermionic field in two dimensions. Additionally, super-renormalizable interactions for standard QFT do not include derivative coupling or counterterms containing derivatives. This means that the commutation relations of the fields are unaffected, and thus, the proof is unaffected [16,17].

To analytically continue in $n$, we ignored the expressions for the $n$ even case to find $S_{\mathrm{out}}$. It would be interesting so see if the expression for all integer $n$ can be analytically continued. One method may be to consider the replicated manifold but for arbitrary real $n$ [i.e., a space with deficit angle $2\pi(1-n)$ at the origin] [21]. Other methods which do not involve analytic continuation could also be used to double check our results [23].

We used the path integral picture to construct our state $\rho(\lambda)$ in Eq. (9) and argue that when evaluating $\mathrm{Tr}[\rho^n]$, one can work on a replicated manifold and interpret the operator $O^{(k)}$ as being inserted on the $(k+1)$th sheet. However, Eq. (9) can be derived without the use of a path integral by noticing that the entanglement entropy of the vacuum state on the distinguished pencil is invariant under translations in the null direction. Furthermore, only the existence of the path integral was required to argue that one can work on the replicated manifold and hence, arrive at Eq. (25). When there are an even number of Majorana spinor fields on the distinguished pencil, one can group them together into Dirac spinor fields, which then can be used to directly construct a path integral. Other methods exists to put even and odd number of Majorana spinor fields into a path integral formalism [24].

There are other proofs of the QNEC. For example, some proofs apply to holographic theories [12,13], while others apply to CFTs with a twist gap [14,15]. After writing this paper, we became aware of a recent general proof of the QNEC by Ceyhan and Faulkner [25], using an approach significantly different from the one in the present paper. Our proof, even though restricted to free fermionic theories, has the advantage of being simpler, gives an explicit expression for $S''_{\mathrm{out}}$, and shows that the proof for the bosonic case [16] can indeed be extended to the fermionic case.

We have implicitly assumed throughout that we are working with the NS (Neveu-Schwarz) sector. We expect a similar proof holds for the R (Ramond) sector.

## ACKNOWLEDGMENTS

We would like to thank Stefan Leichenauer, Tadashi Takayanagi, and Aron C Wall for helpful discussions. We would also like to thank the Department of Physics and Astronomy at The University of Texas at San Antonio and the Yukawa Institute for Theoretical Physics at Kyoto University for financial support for traveling to conferences.

Finally, our thanks to the referee for the constructive comments and for bringing references [14,15,25] to our attention.

## APPENDIX

### 1. Fermionic field

The action for a Majorana (real) fermion in two-dimensional Minkowski space is given by

$$S = k \int dx^2 (-i) \bar{\Psi} \gamma^\mu \partial_\mu \Psi, \quad (A1)$$

where $\{\gamma^\mu, \gamma^\nu\} = 2\eta^{\mu\nu} = 2\mathrm{diag}(1, -1)$, $\bar{\Psi} = \Psi^\dagger \gamma^0$ and $k$ is some normalization parameter, which we will leave unfixed for now. We pick

$$\gamma^0 = \begin{bmatrix} 0 & 1 \\ 1 & 0 \end{bmatrix}, \quad \gamma^1 = \begin{bmatrix} 0 & 1 \\ -1 & 0 \end{bmatrix}, \quad (A2)$$

and so the Majorana condition becomes that the components of $\Psi = \begin{bmatrix} \psi' \\ \bar{\psi}' \end{bmatrix}$ are real. After wick rotation ($t = -i\tau$), this gives us

$$S = -i\frac{k}{2} \int d\tau dx (-i) [\psi' \quad \bar{\psi}'] \gamma^0 (i\gamma^0 \partial_\tau + \gamma^1 \partial_x) \Psi \quad (A3a)$$

$$= -i\frac{k}{2} \int d\tau dx (-i) (i\psi' \partial_\tau \psi' - \psi' \partial_x \psi' + i\bar{\psi}' \partial_\tau \bar{\psi}' + \bar{\psi}' \partial_x \bar{\psi}'). \quad (A3b)$$

Using $iS_E = S$, where $S_E$ is the Euclidean action, we find that

$$S_E = -\frac{k}{2} \int d\tau dx (\psi' \partial_\tau \psi' + i\psi' \partial_x \psi' + \bar{\psi}' \partial_\tau \bar{\psi}' - i\bar{\psi}' \partial_x \bar{\psi}'). \quad (A4)$$

If we define $z = x - i\tau$ and $\bar{z} = x + i\tau$, we find that

$$S_E = -k \int d\tau dx (i) (\psi' \partial_{\bar{z}} \psi' - \bar{\psi}' \partial_z \bar{\psi}'), \quad (A5)$$

To write this action in standard form, we redefine parameters so that

$$S_E = k \int d\tau dx (\psi \partial_{\bar{z}} \psi + \bar{\psi} \partial_z \bar{\psi}), \quad (A6)$$

where $\psi = (-i)^{\frac{1}{2}} \psi' = e^{(\frac{3}{4}\pi i)} \psi'$ and $\bar{\psi} = (i)^{\frac{1}{2}} \bar{\psi}' = e^{(\frac{1}{4}\pi i)} \bar{\psi}'$.

Focusing on the left-moving chiral fermions, we find the reality condition for the redefined parameters is

$$\psi(r)^\dagger = i\psi(r) \quad (A7)$$

for real $r$.[7] Also, since $\psi$ is conformal primary with dimension $(h, \bar{h}) = (\frac{1}{2}, 0)$, we have that

$$\psi(re^{i\theta}) = e^{i\theta/2} e^{\theta k_{\mathrm{pen}}} \psi(r) e^{-\theta k_{\mathrm{pen}}}. \quad (A8)$$

---
[7] $\psi^\dagger(r) = e^{-(\frac{3}{4}\pi i)} \psi' = e^{(-\frac{3}{2}\pi i)} e^{(\frac{3}{4}\pi i)} \psi' = i\psi(r)$

This leads to the reality condition for $f_{\mu\nu}(r, \theta)$ in Sec. III.

After normalization, the two-point correlation function is given by

$$\langle \psi(z) \psi(w) \rangle = \frac{1}{z - w}, \quad (A9)$$

and the chiral (holomorphic) component of the stress tensor is

$$T(z) = -2\pi T_{zz}(z) = -\frac{1}{2} : \psi(z) \partial \psi(z), \quad (A10)$$

where $\partial = \partial_z$. Thus,

$$\langle \psi(z) \psi(w) T(0) \rangle = \frac{1}{2} \left(\frac{1}{zw}\right)^2 (z - w). \quad (A11)$$

The two-point function on the $n$-sheeted replicated manifold for $n$ odd can be found from the 1-sheeted manifold [Eq. (A9)] via a conformal transformation $z \to z^n$ [16],

$$\langle \psi(z) \psi(w) \rangle_n = \frac{(zw)^{\frac{1}{2n}}}{n(zw)^{\frac{1}{2}} (z^{\frac{1}{n}} - w^{\frac{1}{n}})}. \quad (A12)$$

To evaluate,

$$\langle \psi(z-\lambda) \psi(w-\lambda) \rangle''_n |_{\lambda=0} = (\partial_z + \partial_w)^2 \langle \psi(z) \psi(w) \rangle_n, \quad (A13)$$

we first find that

$$(\partial_z + \partial_w) \langle \psi(z) \psi(w) \rangle_n$$
$$= \frac{-(zw)^{\frac{1}{2n}}}{2n^2 (zw)^{\frac{3}{2}} (z^{\frac{1}{n}} - w^{\frac{1}{n}})^2}$$
$$\times [(-1+n) w^{1+\frac{1}{n}} + (1+n) w^{\frac{1}{n}} z - (-1+n) w^{1+\frac{1}{n}}$$
$$- (1+n) w z^{\frac{1}{n}}]. \quad (A14)$$

This expression can be written as

$$\frac{(zw)^{\frac{1}{2n}}}{2(zw)^{\frac{3}{2}} n^2} \sum_{k=0}^{n-1} B_k z^{\frac{k}{n}} w^{\frac{n-k-1}{n}}, \quad (A15)$$

where $B_k$ is to be found by setting Eqs. (A14) and (A15) equal to each other. By doing this, one obtains a recurrence equation for $B_k$,

$$B_k - B_{k-1} = -2 \quad B_0 = n - 1, \quad (A16)$$

which gives $B_k = -2k + n - 1$. Similar methods can be used to rewrite Eq. (A15) as

$$\frac{(zw)^{-\frac{1}{2n}}}{2(zw)^{\frac{3}{2}}n^2} \sum_{k=0}^{n} (k^2 - kn) z^{\frac{k}{n}} w^{\frac{n-k}{n}}. \quad (A17)$$

Similarly, $(\partial_z + \partial_w)^2 \langle \psi(z)\psi(w) \rangle_n$ can be found by applying $(\partial_z + \partial_w)$ from the previous expression to give

$$\langle \psi(z)\psi(w) \rangle_n'' = \frac{-1}{4n(zw)^{\frac{3}{2}}} \sum_{|q| \leq 1} \text{sign}(q)(4q^2 - 1)\left(\frac{w}{z}\right)^q, \quad (A18)$$

where $q \in \frac{\mathbb{Z}}{n} + \frac{1}{2n}$.

### 2. Auxiliary system

By definition,

$$\langle E_{\mu\nu}(\theta) E_{\mu'\nu'}(\theta') \rangle_n = \frac{\text{Tr}[e^{2\pi n K_{\text{aux}}} T[E_{\mu\nu}(\theta) E_{\mu'\nu'}(\theta')]]}{\text{Tr}[e^{2\pi n K_{\text{aux}}}]}. \quad (A19)$$

For $\theta > \theta'$ [16],

$$\text{Tr}[e^{2\pi n K_{\text{aux}}} T[E_{\mu\nu}(\theta) E_{\mu'\nu'}(\theta')]] = e^{-2\pi n K_\mu} e^{(\theta-\theta')\alpha_{\mu\nu}} \delta_{\mu\nu'} \delta_{\nu\mu'}, \quad (A20)$$

and for $\theta < \theta'$ [16],

$$\text{Tr}[e^{2\pi n K_{\text{aux}}} T[E_{\mu\nu}(\theta) E_{\mu'\nu'}(\theta')]] = -e^{-2\pi n K_\mu} e^{(\theta-\theta'+2\pi n)\alpha_{\mu\nu}} \delta_{\mu\nu'} \delta_{\nu\mu'}, \quad (A21)$$

where the extra negative sign comes from the fact that we are dealing with fermionic Grassman-odd operators which anticommute under angle-time ordering.

Considering the case for $\theta > \theta'$. Writing $\theta - \theta' = \Theta$, we can write a Fourier series for

$$e^{\Theta \alpha_{\mu\nu}} = \sum_p e^{-i\Theta(p+\frac{1}{2})} f(p), \quad (A22)$$

where $p$ is summed over all integers divided by $n$. Restricting to $n$ odd and solving for $f(p)$ gives [16]

$$e^{(\theta-\theta')\alpha_{\mu\nu}}$$
$$= -\frac{1}{\pi n} \sum_p e^{-i(\theta-\theta')(p+\frac{1}{2})} \frac{\cosh(n\pi \alpha_{uv})}{i(p+\frac{1}{2}) + \alpha_{uv}} e^{n\pi \alpha_{\mu\nu}}. \quad (A23)$$

This expression is valid for $0 < (\theta - \theta') < 2\pi n$. When $\theta > \theta'$, we can take $\theta - \theta' + 2\pi n$ as our variable in the Fourier series above since $0 < \theta - \theta' + 2\pi n < 2\pi n$ [16] so that

$$- e^{(\theta-\theta'+2\pi n)\alpha_{\mu\nu}}$$
$$= -\frac{1}{\pi n} \sum_p e^{-i(\theta-\theta')(p+\frac{1}{2})} \frac{\cosh(n\pi \alpha_{uv})}{i(p+\frac{1}{2}) + \alpha_{uv}} e^{n\pi \alpha_{\mu\nu}}. \quad (A24)$$

We see that for both the $\theta > \theta'$ and $\theta < \theta'$ case, we can substitute the $e^{(\theta-\theta')\alpha_{\mu\nu}}$ and the $-e^{(\theta-\theta'+2\pi n)\alpha_{\mu\nu}}$ term with the same Fourier series [16]. Hence,

$$\langle E_{\mu\nu}(\theta) E_{\mu'\nu'}(\theta') \rangle_n$$
$$= -\frac{1}{\pi n Z_n^{\tilde{a}ux}} \sum_p e^{-i(\theta-\theta')(p+\frac{1}{2})} \left[\frac{\cosh(n\pi \alpha_{\mu\nu})}{i(p+\frac{1}{2}) + \alpha_{\mu\nu}}\right.$$
$$\left. \times e^{-\pi n(K_\mu + K_\nu)} \delta_{\mu\nu'} \delta_{\mu'\nu}\right], \quad (A25)$$

where $Z_n^{\tilde{a}ux} := \text{Tr}[e^{-2\pi n K_{\text{aux}}}]$ and $Z_1^{\tilde{a}ux} = 1$.

---